\documentstyle[12pt]{article}

\def\[{\left\lbrack}
\def\]{\right\rbrack}
\def\hb{\hfill\break}
\def\({\left(}
\def\){\right)}
\def\{{\left \{}
\def\{{\right \}}
\def\ih{\'\i}
\def\hb{\hfill\break}

\title{NON-POLYNOMIAL LAGRANGIANS IN THE SKYRME MODEL}

\author{Jorge Ananias Neto\\
 Departamento de F\ih sica, ICE \\ Universidade Federal de Juiz de
Fora 36036-330 \\ Juiz de Fora, MG, Brazil }

\date{ }

\begin{document}

\maketitle

\begin{abstract}
\noindent We choose three different coupling constants for a
particular higher-derivative term in the Skyrme model that allows the total
Lagrangian to converge in a  binomial,
geometric and a logarithmic form. Improved numerical results are obtained.
\end{abstract}

\vskip 1 cm


\newpage

\noindent The Skyrme model\cite{Skyrme} consists of treating baryons
as soliton solutions in the non-linear Sigma model with an additional
stabilizer Skyrme term. We have recently studied \cite{J A Neto}
the introduction of another four derivative stabilizer term given by

\begin{equation}
\label{l2g}
 L_2 = \int d^3r \, c_2 \[ Tr \( \partial_\mu U
\partial^\mu U^+ \) \]^2 \,,
\end{equation}

\noindent where $c_2\equiv {1\over 32 e^2}\,$. This form, known as the
symmetric Skyrme term, is the
square of the chiral SU(2) Sigma Model whose original Lagrangian is

\begin{equation}
\label{l1}
L_1={F^2_\pi\over16} \int d^3r \, Tr\left(\partial_\mu U \partial^\mu U^+
 \) \,,
\end{equation}

\noindent where $\,F_\pi\,$ is the pion decay constant. Using the
stabilizer $L_2$ term as a pattern for the inclusion of the
higher-order derivative
terms, we have verified an improvement in the physical values given by
Skyrme Model. The purpose of this paper is to perform a study
about the possibility of choosing different parameter relations in
the higher derivative terms, consequently leading to a new types
of Skyrme-like Lagrangian. \par Thus, the standard form of the
Lagrangian terms is

\begin{equation}
\label{ln}
 L_n = \int d^3r \, c_n \[Tr\( \partial_\mu U \partial^\mu U^+ \) \]^n \,,
\end{equation}

\begin{sloppypar}
\noindent where n=1,2,... . Performing the usual collective
coordinate expansion\cite{ANW} $\,U(r,t)=A(t)U_0(r)A^+(t),\,$
being A  a SU(2) matrix\footnote{ Consequently, the matrix A can be written as
$ A=a_0+ i \tau_j a_j $. }, we can write the Lagrangian
(\ref{ln}) in the form
\end{sloppypar}

\begin{equation}
\label{lnc}
L_n=\int d^3r \, c_n \[ -2M+I \,Sp \]^n \,,
\end{equation}

\noindent where M is given by M\footnote{Here we have used
the hedgehog ansatz, $\,\, U={\exp \(i \tau . \hat r F(r)\)} \,$,
 where F(r) is called the chiral angle.} $\equiv  \[{2 \sin ^2F(r)\over r^2}
+ F'^2(r) \], \,$  the inertia moment I is written as
$I\equiv {8\over3} \sin^2F \,$ and Sp is given by
$Sp\equiv Tr\[ \partial_0 A \partial_0 A^{-1} \].\,$ We can read the
final form of the Lagrangian containing all derivative terms as

\begin{eqnarray}
\label{lns}
  &L= -c_1 \int d^3r \,\[ 2M-I\,Sp \]
 -c_2 \int d^3r \,\[ 2M-I\,Sp \]^2  \nonumber \\
 & \dots -c_n \int d^3r \,\[ 2M-I\,Sp \]^n \, .
\end{eqnarray}

\noindent This arrangement, as we will see in the next sections, ensures
the positivity of the total Hamiltonian. Then, the idea of
this work consists in choosing different parameter relations
that permit to sum the Lagrangian with higher derivative terms
in a well known specific form. We set three different coefficient
relations \cite{grad}: a) $ K_n\equiv {c_n\over c_1}=  \left( \begin{array}{c}
s \\ {n-1} \end{array} \right)/(2e^2F_\pi^2)^{n-1},\,\, n=1,2,3 \dots\,, $
which permits
the total Lagrangian to converge to a binomial form;
 b) $ K_n\equiv {c_n\over c_1}= {1\over (2e^2
F_\pi^2)^{n-1}}\,\,, n=1,2,3 \dots , $ in which the total
 Lagrangian is summed in a geometric series form; and
c)$  K_n\equiv {c_n\over c_2}= {(-1)^{n-1}\over
(n-2) (2e^2 F_\pi^2)^{n-2}}\,\, ,\,  n=3,4,5 \dots \,,$
which permits the total Lagrangian to converge to a logarithmic
form.

\vskip 1.5 cm

\noindent {\large {\bf  Binomial Form}}

\vskip .8 cm

\noindent If we set $ K_n\equiv {c_n\over c_1}=
\left( \begin{array}{c} s \\ {n-1} \end{array} \right) /
(2e^2F_\pi^2)^{n-1}\,\,$
the total Lagrangian (\ref{lns}) converges in a binomial form
when $ n \rightarrow \infty \, ,$

\begin{eqnarray}
\label{bin}
   &L=- c_1 \int d^3r \, \[ 2M-I\,Sp \] [ 1 +
{ c_2 \over c_1} \int d^3r \, \[ 2M-I\,Sp \]
\nonumber \\ &  \dots  +{c_n\over c_1} \int d^3r \, \[ 2M-I\,Sp \]^{n-1} ]
\nonumber \\\ & = -c_1 \int d^3r [ 2M-I \, Sp ] [ 1 + {2M-I \, Sp \over 2e^2
F_\pi^2} ]^s\,,
\end{eqnarray}
\begin{sloppypar}
\noindent where $ c_1\equiv{ F_\pi^2 \over 16} \,$.
In the process of collective coordinates quantization it is
necessary that in the Lagrangian (\ref{bin}) we have only a
linear term in {\it Sp}. Then, making a Taylor series expansion
and retaining only the linear
term in {\it Sp}, we obtain an expression for the Hamiltonian
\footnote{ The Hamiltonian H is defined by $ H= \pi_i \dot a_i - L \,,$ where $
\pi_i={ \partial L \over
\partial \dot a_i }. $ },
written as \end{sloppypar}

\begin{equation}
\label{HQOP}
H= M_T+{1\over 8I_T} \sum_{i=0}^3 \pi_i^2 \,\,  ,
\end{equation}

\noindent where

\begin{equation}
\label{MTB}
M_T = {F_\pi \over {\it e}} {\pi \over 2} \int_0^\infty dx x^2 M \[ 1+M \]^s
\,,
\end{equation}

\noindent and

\begin{eqnarray}
\label{inerb}
I_T= {2\pi \over 3} {1\over e^3F_\pi} \int_0^\infty dx x^2 \sin^2F
\[ s \, M \( 1+M \)^{s-1} + \( 1+M \)^s \] \,.
\end{eqnarray}

\noindent In the last expression we have used the dimensionless variable
$x={\it e} F_{\pi}r$. \par The  quantized Hamiltonian
form is obtained taking $ \,\pi_i=-i {\partial \over \partial a_i} \,$,
which leads to

\begin{equation}
\label{HQEB}
H= M_T + {1\over 8I_T} \(-{\partial^2\over \partial a_i^2}\)=
M_T+{l(l+2)\over 8I_T}\,, \,\,\, \, l=1,2,3,\dots \,\, .
\end{equation}

\begin{figure}
\vspace{9.cm}
\caption[]{Behavior of the parameter B defined by
$B\equiv x^2F(x)\,,$ where F(x) is the numerical variational
solution of the classical binomial Hamiltonian form including
terms up to power s=3/2, s=5/2, s=7/2, s=11/2, and s=21/2.}
\label{binomio}
\end{figure}

\begin{sloppypar}
\noindent The numerical solution of F(x) with the boundary conditions
$ \, F(0)=\pi \,$ and $ \, F(\infty)=0 \,,$ for different values of s powers,
is obtained using the variational Euler-lagrange equation which is given by

\newpage

\begin{eqnarray}
\label{Eulerb}
 & [ 8x^2F'^2 s W^{(s-1)} + 4 x^2 F'^2 s (s-1) (W-1) W^{(s-2)}+
2 x^2 W^s\nonumber \\ & + 2 x^2 s (W-1) W^{(s-1)} ] F''+4xW^s F'+2 x^2 F' s
W^{s-1} dax\nonumber \\ & +4x F' (W-1) s W^{(s-1)}+2 x^2 F's W^{(s-1)} dax
\nonumber \\ &+2 x^2 F' s (s-1)(W-1) {W^{(s-2)}} dax\nonumber \\ & -x^2 W^s daf
- x^2 (W-1) s {W^{(s-1)}} daf=0 \,\,\,,
\end{eqnarray}
\end{sloppypar}

\noindent where $W \equiv [1+{2\sin^2 F \over x^2}+F'^2 ] \,$,
dax $\equiv [  {2\sin 2F F'\over x^2} - {4 \sin^2 F \over x^3} ] \, $
and daf $\equiv [ {2 \sin 2F\over x^2} ] \, .$
\vskip .2 cm

\noindent The soliton solution
is obtained if we impose the condition,$ \,\lim_{x\rightarrow \infty}
F(x) = {B\over x^2} \,.$ In figure 1, we show the numerical
behavior of parameter B, which is directly proportional to the axial
vector coupling constant, $\,g_A={2\pi\over 3} {B\over e^2}\,$.

\par Using the masses
of the Nucleon   $ (M_N=939 Mev)$ and of the Delta $ (M_\bigtriangleup=
1232 Mev)$ as input parameters, we determine the pion decay constant
 $ F_ \pi $
and the dimensionless Skyrme parameter $ {\it e}.\,\,$
\noindent The main physical results are shown in table 1,
 according to ref.\cite{ANW},

\vskip .7 cm

\centerline{TABLE 1- Physical parameters in the Skyrme Model}
\centerline{ (binomial form)}

$$\vbox{\halign{\hfil#\hfil&\quad\hfil#\hfil&\quad\hfil#\hfil&\quad\hfil#\hfil&
\quad\hfil#\hfil&\quad\hfil#\hfil&\quad\hfil#\hfil&\quad\hfil#\hfil&
\quad\hfil#\hfil\cr
\noalign{\hrule}
\ \cr
s &3/2 & 5/2 & 7/2 & 11/2& 21/2& ANW & expt. \cr
\ \cr
\noalign{\hrule}
\ \cr

{F$_\pi(Mev)$} & 136 & 140 & 142 & 143 & 143 & 129 & 186 \cr

{\it e} & 6.81& 9.65 & 11.81 & 15.29 & 21.55& 5.45 &- \cr\cr

$<r^2>^{1\over2}_{I=0}(fm)$ & 0.60 & 0.60 & 0.61 &0.61 &0.61&0.59 & 0.72\cr

$\mu_p$ & 1.74 &  1.74 & 1.74 &1.74 & 1.75 & 1.87 & 2.79\cr

$\mu_n$ & -1.22 & -1.21 & -1.21 & -1.21 &-1.20 &-1.33 &  -1.91\cr

$g_A$ & 0.73 & 0.76 & 0.77 & 0.78 & 0.79 & 0.61 & 1.23\cr

\ \cr
\noalign{\hrule}}}$$

\vskip .7 cm

\noindent where ANW are the results of Adkins, Nappi and Witten (ref.
\cite{ANW}). These results indicate a convergence of the physical parameters
to stable values with the increase of parameter {\it s}.

\vskip 1.5 cm

\noindent{\Large{\bf  Geometric Form}}

\vskip .8 cm

\noindent If we set $ K_n\equiv {c_n\over c_1}= {1\over (2e^2
F_\pi^2)^{n-1}}\,\, $ the total Lagrangian (\ref{lns}) converges
in a geometric form when $ n \rightarrow \infty \, ,$

\begin{eqnarray}
\label{geo}
&L=- c_1 \int d^3r \, \[ 2M-I\,Sp \] [ 1 +
{ c_2 \over c_1} \int d^3r \, \[ 2M-I\,Sp \]
\nonumber \\ &  \dots  +{c_n\over c_1} \int d^3r \, \[ 2M-I\,Sp \]^{n-1} ]
\nonumber \\\ & = -c_1 \int d^3r \[ 2M-I \, Sp \] \[  1-{ \[2M-I \, Sp \]
\over 2e^2 F_\pi^2} \]^{-1}\, ,
\end{eqnarray}

\vskip  .9 cm

\noindent where $ c_1\equiv{ F_\pi^2 \over 16} \,$. Using the same procedure
adopted in the binomial section, i.e.,
retaining only the linear term {\it Sp} in the Lagrangian (\ref{geo}),
we can write the eigenvalues of the quantized Hamiltonian as

\begin{equation}
\label{HQG}
H= M_T+{l(l+2)\over 8I_T}\,, \,\,\, \, l=1,2,3 \dots ,
\end{equation}

\noindent where

\begin{equation}
\label{MTG}
M_T = {F_\pi \over {\it e}} {\pi \over 2} \int_0^\infty dx x^2 M
\[ 1- M \]^{-1} \,,
\end{equation}

\noindent and

\begin{eqnarray}
\label{inerg}
I_T= {2\pi \over 3} {1\over e^3F_\pi} \int_0^\infty dx x^2 \sin^2F
\[ M \( 1-M \) ^{-2} + \( 1-M \) ^{-1} \] \,\, .
\end{eqnarray}

\noindent The Euler-Lagrange equation which gives the numerical
solution of F(x) with the boundary conditions $ \, F(0)=\pi \,$ and $ \,
F(\infty)=0 \,,$ is given by,

\begin{eqnarray}
\label{Eulerg}
\[ 2 x^2+ 8 x^2 F'^2 \, {Wg}^{-1} \] F^{\prime\prime}
+ 4 x F' + 4 x^2 F' {Wg}^{-1} dax \nonumber \\  - x^2 daf =0 \,\,\,,
\end{eqnarray}

\noindent where $Wg \equiv [1-({2\sin^2 F \over x^2}+F'^2) ] \,$, and
dax and daf are defined in (\ref{Eulerb}). In
figure 2 we show the behavior of the soliton parameter B, and
as we have observed in the binomial case, this parameter is directly
proportional to the axial
vector coupling constant, $\,\, g_A={2\pi\over 3} {B\over e^2}\,$.

\begin{figure}
\vspace{7.5cm}
\caption[]{Behavior of the parameter B defined by
$B\equiv x^2F(x)\,,$ where F(x) is the numerical variational
solution of the classical geometric series Hamiltonian form.}
\label{geometrico}
\end{figure}

Again, using the masses of the Nucleon  and of the Delta as input parameters,
we obtain the main physical results, which are shown in table 2.

\vskip .5 cm

\centerline{TABLE 2- Physical parameters in the Skyrme Model}

$$\vbox{\halign{\hfil#\hfil&\quad\hfil#\hfil&\quad\hfil#\hfil&\quad\hfil#\hfil
\cr
\noalign{\hrule}
\ \cr
  & geometric form & ANW & expt. \cr
\ \cr
\noalign{\hrule}
\ \cr

{F$_\pi(Mev)$} & 152 & 129 & 186 \cr

{\it e} & 8.48 & 5.45 &- \cr\cr

$<r^2>^{1\over2}_{I=0}(fm)$ & 0.61&0.59 & 0.72\cr

$\mu_p$  & 1.75 & 1.87 & 2.79\cr

$\mu_n$  &-1.21 &-1.33 &  -1.91\cr

$g_A$ & 0.84 & 0.61 & 1.23\cr

\ \cr
\noalign{\hrule}}}$$
\vskip .7 cm

\vskip 1.5 cm

\noindent {\large {\bf  Logarithmic Form}}

\vskip 1 cm

\noindent Defining $ K_n\equiv {c_n\over c_2}=
{(-1)^{n-1}\over
(n-2) (2e^2 F_\pi^2)^{n-2}}\,\, , n=3,4 \dots \,\,$,
the total Lagrangian (\ref{lns}) converges in a logarithmic specific form
which admits a soliton solution when $ n \rightarrow \infty \,, $

\begin{eqnarray}
\label{log}
&L=- c_1 \int d^3r \, \[ 2M-I \, Sp \]- c_2 \int d^3r \, \[ 2M-I \, Sp \]^2
[ 1 \nonumber \\ & +{ c_3 \over c_2} \int d^3r \, \[ 2M-I \, Sp \]
 \dots  +{c_n\over c_2} \int d^3r \, \[ 2M-I \, Sp \] ^{n-1} ] \nonumber \\\
& = -c_1 \int d^3r \[ 2M-I \,Sp \]\nonumber \\ & - c_2 \int d^3r \[2M-I \, Sp
\] ^2 \[ 1 + \log \( 1+{
2M-I \, Sp\over (2e^2 F_\pi^2)} \) \]\,,
\end{eqnarray}

\noindent where $ c_1\equiv{ F_\pi^2 \over 16} \,$ and $ c_2 \equiv {1\over
{32 e^2}} \, . \, $ The total quantized eigenvalues Hamiltonian are given by

\begin{equation}
\label{HQEL}
H= M_T+{l(l+2)\over 8I_T}\,, \,\,\, \, l=1,2,3\dots ,
\end{equation}

\noindent where

\begin{equation}
\label{MTL}
M_T = {F_\pi \over {\it e}} {\pi \over 2} \int_0^\infty dx x^2 M \[ 1
+M \( 1 + \log \( 1+ M \) \)\] \,,
\end{equation}

\noindent and

\begin{eqnarray}
\label{inerl}
I_T= {2\pi \over 3} {1\over e^3F_\pi} \int_0^\infty dx x^2 \sin^2F \,\,
[ \, 1 + M \,( M (1+M )^{-1} \nonumber \\ + 2 ( 1 + \log ( 1 + M )\, ) ) \,]
\,\, .
\end{eqnarray}

\noindent The Euler-Lagrange equation which gives the soliton solution
with the boundary conditions $ \, F(0)=\pi \,$ and $ \, F(\infty)=0 \,,$ is

\begin{eqnarray}
\label{Eulerl}
 &[2 x^2+4x^2M+8x^2F'^2+4x^2F'^2 Sl+2x^2M Sl +12x^2F'^2 M (1+M)^{-1}
\nonumber \\ & -4x^2F'^2 M^2 (1+M)^{-2}] F^{\prime\prime}+ 4xF'+8xF'M+4x^2F'
dax +4xF' M Sl\nonumber \\ &+2x^2F' dax Sl + 6x^2F' M (1+M)^{-1} dax -2x^2F'
M^2 (1+M)^{-2} dax\nonumber \\ &-2 \sin(2F) -2x^2 M daf -2x^2 M daf \log(1+M)
\nonumber \\ & - x^2 M^2 (1+M)^{-1} daf =0 \,\,\,,
\end{eqnarray}

\begin{figure}
\vspace{7.5cm}
\caption[]{Behavior of the parameter B defined by
$B\equiv x^2F(x)\,,$ where F(x) is the numerical variational
solution of the classical logarithmic series Hamiltonian form.}
\label{logaritmico}
\end{figure}

\noindent where $Sl\equiv \[ 2 \log(1+M) + M(1+M)^{-1} \]$, and dax and daf
 are defined in (\ref{Eulerb}). With the masses of the Nucleon and of the Delta
as input parameters, the physical results (table 3) are given by

\vskip .7 cm

\centerline{TABLE 3- Physical parameters in the Skyrme Model}

$$\vbox{\halign{\hfil#\hfil&\quad\hfil#\hfil&\quad\hfil#\hfil&\quad\hfil#\hfil
\cr
\noalign{\hrule}
\ \cr
  & logarithmic form & ANW & expt. \cr
\ \cr
\noalign{\hrule}
\ \cr

{F$_\pi(Mev)$} & 141 & 129 & 186 \cr

{\it e} & 6.69 & 5.45 &- \cr\cr

$<r^2>^{1\over2}_{I=0}(fm)$ & 0.60 & 0.59 & 0.72\cr

$\mu_p$  & 1.74 & 1.87 & 2.79\cr

$\mu_n$  &-1.21 &-1.33 &  -1.91\cr

$g_A$ & 0.76 & 0.61 & 1.23\cr

\ \cr
\noalign{\hrule}}}$$
\vskip .7 cm

\par Before going to the final comments, we would like
to mention that there are many controversies about the possibility
of Lagrangian terms which contain quartic or more time derivatives
destabilizing the Skyrmion \cite{Kal}. At first, as we have described
in the binomial section, this problem can be overcome with the linear
procedure used in the collective coordinates expansion. Then, it is also
interesting to study the process of quantization of collective
coordinates including  higher derivative time terms.
This question is a natural step of continuation of this work and
it will be object of a forthcoming paper.
\par The three Skyrme-type Lagrangians that are presented in
this paper; that is, the binomial, geometric and logarithmic plus exponential
form developed in \cite{J A Neto} are an
attempt to include, in a simple way, the contribution of a specific
higher derivative term(\ref{ln}). We observe that the physical
parameters which result from this procedure go in the right
direction. It is interesting to point out that our binomial Skyrme
form resembles the unconventional Born-Infeld electrodynamics
worked out by Dirac \cite{Dirac}.
Here we would like to mention that the best results are obtained
using the geometric form. We hope that with the
four different forms of the Skyrme Lagrangian we have covered
the main contributions of particular higher derivative terms( known as
the symmetric Skyrme term)
in the usual physical values of the Skyrme model.
\par I would like to thank M.G. do Amaral for critical reading and
the \break Departamento de Campos e Part\ih culas of Centro Brasileiro
de Pesquisas F\ih sicas for hospitality while part of this work
was carried out.

\end{document}